\documentclass[letterpaper,twocolumn,10pt]{article}
\usepackage{usenix,epsfig,endnotes}
\usepackage{pdfcomment}
\PassOptionsToPackage{hyphens}{url}\usepackage{hyperref}
\usepackage{breakurl}

\begin{document}

\date{}

\title{ \bf Academic Cloud Computing Research: Five Pitfalls and Five Opportunities
}
\author{
{\rm Adam Barker, Blesson Varghese, Jonathan Stuart Ward and Ian Sommerville}
\\ School of Computer Science, University of St Andrews, UK
}

\maketitle

\subsection*{Abstract}
This discussion paper argues that there are five fundamental pitfalls, which can restrict academics from conducting cloud computing research at the infrastructure level, which is currently where the vast majority of academic research lies. Instead academics should be conducting higher risk research, in order to gain understanding and open up entirely new areas. 

We call for a renewed mindset and argue that academic research should focus less upon physical infrastructure and embrace the abstractions provided by clouds through five opportunities: user driven research, new programming models, PaaS environments, and improved tools to support elasticity and large-scale debugging. The objective of this paper is to foster discussion, and to define a roadmap forward, which will allow academia to make longer-term impacts to the cloud computing community.

\section{Introduction}

Imagine a world where there was no academic cloud computing research. Would the field of cloud computing really be any different from what it is today? Would it have evolved in the same way? This paper argues that academics are addressing the wrong class of problems to have any lasting impact, and that cloud computing has evolved without substantial influence from academia. 

The core of this problem comes down to scale: academics generally do not have low-level infrastructure (e.g., compute, storage, network) access to data centre sized resources, in order to design, deploy and test algorithms. To circumvent this access problem academics simulate the behaviour of data centres, and run experiments on small (toy) private clouds deployments. If this research is then deployed at scale, there is the ever present risk it is non representative of real world data centres, and is therefore of limited value to the community. In addition, a large proportion of academic research focuses on incremental improvements to existing frameworks created by industry, e.g., Hadoop optimisation for a small set of edge cases. Typically these contributions are short-lived and have no lasting impact on the cloud computing community; not because these contributions are irrelevant, but  because these kinds of problems are already being solved by industry, who do have access to low-level infrastructure at scale \cite{welsh}. 

Instead academics should be conducting higher risk research, which sheds light on more complex problems, in order to gain understanding and open up entirely new areas. We expect to see genuine cloud research focus less upon physical infrastructure and continue to embrace abstraction through user driven research, programming models, and Platform as a Service (PaaS) clouds. In addition, research needs to focus on providing software engineers with better tools to program truly elastic applications, and debug large-scale applications, which utilise existing cloud infrastructure. The objective of this paper is to foster discussion, and to define a roadmap forward, which will allow academia to make longer-term impacts. Many of these observations have been derived from an eight year UK research programme focused on Large Scale Complex IT Systems (LSCITS) \cite{LSCITS}.  

\section{Five Pitfalls of Academic  Research}

\subsection*{Pitfall 1: Infrastructure at Scale}

The presence of unused capacity within large-scale data centres led to the development of what became known as cloud computing. Therefore, since its inception, cloud computing and scale are inseparably linked. Today, rapid scalability and on demand resource provisioning are afforded to users through the ever increasing scale of cloud providers. Current estimates of the scale of Amazon EC2, the dominant public cloud provider, vary between 156,225 and 454,400 servers~\cite{netcraft}. Amazon Web Services (AWS) are now the largest web host on the planet, which hosts numerous high profile customers including Netflix, Instagram, DuckDuckGo and Reddit. 

The scale of AWS and equivalent cloud providers (in terms of servers, users, data, etc.) cannot be easily replicated. To produce a similar environment requires significant capital expenditure, which is extremely prohibitive for academic researchers. At best, academic researchers can hope to reproduce a tiny subsection of the resources available through AWS. Typically this is achieved though the use of a small testbed built from commodity hardware running, for example OpenStack~\cite{openstack}. Such deployments typically number between the tens and hundreds of servers, supported by gigabit Ethernet and commodity network attached storage. As such they cannot come close to replicating the scale of the compute, network or storage capacity that is expected from even the smallest public clouds.

In order to perform an evaluation which indicates performance on a public cloud,
there is no substitute for a real world public cloud deployment. Researchers
who wish to investigate applications or phenomena running on top of Virtual Machines (VMs) and storage from a public cloud provider can feasibly do so. Any research which must be evaluated
at scale can incur significant costs but remains a feasible means of evaluation. 

If however research is focused upon low-level compute, storage, network, energy or other systems there is the ever present risk of producing an evaluation which is non representative of
real world cloud deployments. Without access, academics
cannot modify and evaluate low-level cloud mechanics, making useful research in this area
all but infeasible. Work which is targeted at the cloud infrastructure
level and evaluated on a non representative private testbed can at best 
deliver interim results, which suggest how it may behave when deployed in a
real world context.

This has caused a dichotomy within academic cloud research: between those researchers with 
access and partnership with cloud providers and those without. Two recent examples to illustrate the point include: the latest best paper award at IEEE Cloud 2013 \cite{nec} was awarded for collaborative research between Princeton and NEC, the best paper at SIGCHI (non-cloud research but requires access to data) was awarded for research between Facebook and Carnegie Mellon University \cite{facebook}. This access problem prevents academics from having any lasting impact, and  has significant implications for systems research investigating the lower levels of the cloud stack. For academics that are used to low-level access this barrier to research is unprecedented. 

%

There are efforts underway with the European Union BONFIRE \cite{BONFIRE} project, and Open Cirrus \cite{opencirrus} to provide academics with access to large-scale resources. Individual academics are predominantly concerned with research and not maintaining testbeds and research systems. A high turnover of researchers and a emphasis on research (and not maintenance) often leads to the neglect and eventually the premature abandonment of testbeds. To ensure the longevity of a research system, with appropriate documentation and maintenance, dedicated staff are necessary. This cannot be easily afforded be individual institutions. Academics need to collaborate, self organise and build their own data centres at the national level (through government funding etc.) in order to provide the facilities and maintenance necessary for academics to conduct meaningful cloud computing research. 


\subsection*{Pitfall 2: Abstraction}
Abstraction is one of the key features of the cloud computing model. The main principle of public cloud computing models is that a user can make use of VMs as if they were offered as a service, and the complexities of managing and maintaining hardware are abstracted from an end user. This translates into the fact that such an abstraction results in an `observation-only' or a `black box' model. Unlike the grid, the low-level machine details are concealed on the cloud. This has a number of compelling advantages over grid computing architectures including elastic on-demand architecture and high service availability \cite{cloudcompared-2}.

However, in academic cloud computing research, such an abstraction in which the cloud is merely used as a service or tool for computing has been of minimal interest, e.g., deployment of a Physic's application on the cloud to compare performance with a cluster environment. In addition, a lot of academic cloud computing research has a reverse engineering focus, where low-level infrastructure components are often (poorly) reimplemented for the purposes of research prototypes. This not only contradicts the underlying principle of public cloud computing models (to be used as a service), but such research meets with little support from the cloud providers themselves. However, research which has built on top of existing cloud platforms in order to provide new functionality has been enormously successful. RightScale \cite{rightscale} is an exemplar of how academic cloud computing research has added innovative new features and been commercialised. 

\subsection*{Pitfall 3: Non-reproducible Results} 

As academics do not have low-level access to data centre resources, why not simulate a data centre instead? This is a reasonable response from academia to the problem of access to such large-scale resources, and has worked well in the networking community, e.g., NS3 \cite{NS3}.  There are multiple simulators, which attempt to model and simulate a cloud computing environment, including CloudSim \cite{cloudsim} and GreenCloud \cite{greencloud}. Simulations can be effectively used as a prototyping mechanism to provide a rough idea of how a particular algorithm may perform. The problem with each of these simulation tools is that it is very difficult to verify if the simulation environment is an accurate representation of a real world data centre environment. To date there have been no attempts by the community to align simulations with real world data from industry -- refer to pitfall 5 for a further discussion of industrial relations. Furthermore, real world data centres are constantly evolving, and are subject to both planned and unplanned changes. Therefore even if a simulation model is verified at a particular point in time, this will no longer hold if the simulation is rerun at a later point in time. Research which is built on top of these simulation environments (e.g., simulation of VM migration) has next to no real world application, other than to demonstrate a prototype algorithm or application. 

An alternative approach for academics without direct access to large-scale infrastructure is to build simulations with real world trace data. While there are a few examples \cite{Benson, SWIM} of companies releasing trace data for academics, we consider this to be the exception rather than the rule, and traces are only available for a very limited set of applications. 

In order to get a paper accepted to one of the top-tier cloud computing conferences a thorough evaluation must be presented for peer-review. Usually this evaluation has a very specific environment in terms of hardware and software, which may very well be a bespoke setup up for the purposes of getting a paper accepted for publication. Reproducing these results is next to impossible unless a replica of the evaluation environment can be deployed, or access is granted to the original environment. Furthermore, results of an evaluation are usually not general purpose; just because an expected behaviour was observed on a 10 node private cloud, does not mean that the same results will be observed on a 10,000 node data centre. There are currently no reliable mathematical models of how clouds scale over time, this is something that we highlight as an immediate research opportunity, and can only be solved through improved relationships between academia and industry, discussed further in pitfall 5. 

\subsection*{Pitfall 4: Rebranding}

Academia is notoriously trend driven, and in order to gain competitive funding, government PhD programmes and collaboration academics must be working in certain current areas of research. Much of academic cloud computing research is simply grid computing, or cluster computing, or e-Science rebranded as cloud computing research. 

The novelty of the cloud over preceding infrastructures such as clusters and grids include features such as elasticity, scalability, on-demand availability and minimal user maintenance of the resources \cite{cloudcompared-1}. The majority of academic cloud computing research does not require any of these core properties, which differentiate clouds from other forms of infrastructure. 


Researchers within the cloud domain must embrace the avenues of research which the paradigm allows for. Research which places its focus at the bottom of the cloud stack at the level of virtualisation, OS and systems research may have significant fundamental implications for cloud computing, but many of these results cannot be tested by the wider academic community due to lack of access to infrastructure at scale listed in pitfall 1. We believe that academic research which operates higher up the stack is likely to have a longer-term impact.

\subsection*{Pitfall 5: Industrial Relations}

As discussed throughout this short paper, the only effective way to develop and test new cloud services is to deploy them at a scale. Unfortunately there are virtually no mechanisms which allow academics to gain access to data centres in order to conduct low-level systems research. This is a completely understandable stance by industry, why would a company like Google, or Amazon allow academics access to production, or even test data centres in order to evaluate new academic research? Visiting programmes at Google such as the Visiting Faculty Programme \cite{google} do exist, but are not available to the vast majority of the academics working on cloud computing research. There are examples of University research labs working in tandem with cloud infrastructure providers; a recent success story is the Algorithms Machines and People Lab (AMPLab) \cite{amplab}, at UC Berkeley. However, these relationships are difficult to establish, and are only available to the very top-tier institutions with access to reliable industry contacts.

There are multiple programmes which allow academics to use the cloud as a service, for example the AWS Education programme \cite{aws}, Windows Azure for Research programme \cite{azure}, and Google App Engine Research Awards \cite{app_engine}. These programmes are ideal for building and testing applications which sit within an IaaS or PaaS cloud. However these programmes do not solve the problem with low-level infrastructure access, which is where the majority of current academic research lies. 

\section{Five Opportunities for Academic Research}


\subsection*{Opportunity 1: User Driven Research}
We believe the enthusiasm in the academic community for cloud research needs to be directed towards problems that genuinely require the exploitation of elasticity, scalability and abstractions offered by the cloud. A number of such problems exist in science, but a large proportion can be addressed on clusters and simply do not require the core properties of clouds. The requirements of such domain experts need to be understood and then cloud-based environments developed in order to facilitate scientific discovery. Here we make the differentiation between building clouds for science applications, which is not what we are suggesting, but using existing cloud services effectively, which make use of elasticity, scalability and suitable abstractions. One potential area that can be explored in this context are virtual private research environments using simple cloud tools but tailored to suit specific domain requirements. Such an approach is adopted in the ELVIRA \cite{elvira} and CARMEN \cite{carmen} projects.

Another area that needs to be investigated is the development of environments that support budget-limited computation based on a set of user driven requirements. Let us assume that a job needs to be executed for a prolonged period but is constrained in the total available budget for computation. How can the job be most effectively executed across clouds, and what models of predicting computational workloads and calculating costs need to be developed. Here again we make a differentiation between developing cost models for the cloud, and cost computation and prediction models on top of existing clouds, and the latter is our suggestion for pursuing in an academic environment. User driven research on the cloud overlaps with research that needs to focus on PaaS environments which is considered in opportunity 4. 

\subsection*{Opportunity 2: Programming Models}

Engineers need more efficient, scalable and usable programming models, which will allow them to take full advantage of the core properties of cloud computing. Through the use of programming models built on top of cloud resources, developers should be able to write scalable, elastic code without having to mange low-level cloud primitives. 

At present, MapReduce is the default programming model for applications which deal with large data sets, and has been a highly usable approach to dealing with data. However, MapReduce is not always the appropriate model and there are seldom alternatives. There is some variation available from tools which are built upon MapReduce including Apache Hive \cite{hive}, Cloudera Impala~\cite{impala} and Apache Pig \cite{pig}. These tools offer programming abstractions which are not available in vanilla Hadoop, however they still rely upon the underling MapReduce functionality.

There are several alternative programming models which are beginning to emerge which deviate from conventional MapReduce semantics. The Apache Spark~\cite{spark} project is a successful example of a project which has made the transition from academic research into production cluster environments in major companies. Spark is an engine for large-scale data processing which offers higher levels of performance for certain application classes when compared to MapReduce.
Another emerging programming model is Apache Drill \cite{drill}, an open source reimplementation of Google BigQuery which is designed for high throughput queries on very large datasets. Drill is currently an incubator project at Apache and like many other programming models hinges on further development and mass adoption to ensure its success. 

\subsection*{Opportunity 3: Bugs in Large-Scale Applications} 

Large-scale systems are inherently difficult to manage. The level of complexity
and rate of change in such systems presents a constantly moving target, which
requires substantial talent and manpower to mange. Debugging large-scale cloud applications is exceptionally difficult, due to
the scale of a deployment, the abstraction from the underlying hardware, and
the nature of a remote deployment. Research should focus on a tool chain which allows
engineers to collect, visualise and inspect the state of jobs running across
many thousands of cores.  Tools should focus on the emergent properties of large-scale applications; properties which can only be observed when applications are deployed at the scale offered by cloud resources. Research into this area can be built on top of existing cloud platforms and wouldn't necessarily require low-level access to large-scale testbeds. 

In academia, however, this area has received 
insufficient attention hitherto the means to develop and debug large-scale applications
are under researched. There remains only a small body of research in this crucial
area~\cite{Zhou:2013:WAD:2493123.2462907, Chilimbi:2009:HES:1555001.1555020}, which limits the ability of academics to translate prototypes into real world
software. A lack of knowledge forces current research to be discarded
or significantly rewritten in order to be applicable to large-scale systems. Through
further investigation into the software engineering of large-scale systems, academics 
can produce superior cloud research which supports the scale of systems that are
now becoming commonplace.

\subsection*{Opportunity 4: PaaS Environments}
Platform as a Service (PaaS) clouds aim to offer a higher level of abstraction, away from bare-bones infrastructure services. At this level of the cloud stack the focus is on automating certain tasks, which if done by manually writing programs can be a time consuming task. In other words an abstract environment, which incorporates elasticity and on-demand service features offered through an easy to use interface \cite{paas-1, paas-3}.  


Generic frameworks such as StarCluster \cite{StarCluster} and Cloud Foundry \cite{cloudfoundry} exist to address some of the automation described above. PaaS environments for science are gaining momentum (for example, Helix Nebula: The Science Cloud~\cite{helix}). We believe this is one avenue of research that can benefit from being pursued. Here the focus needs to be placed on building environments, which make use of cloud infrastructure provided by multiple vendors, and at the same time providing a high-level interface that will allow scientists to quickly and easily access the resources. 
Such research can reap maximum benefit if based on user driven requirements and directed towards developing stand-alone and independent environments, which make use of the variety of PaaS features offered by the providers.

\subsection*{Opportunity 5: Elasticity}
The most significant benefits of cloud computing are obtained by applications that scale over a large number of resources. Traditionally, in grid and cluster environments the resources requested for an application had to be provisioned before its execution. Cloud computing, however, facilitates dynamic resource provisioning and provides a high level abstract interface for doing this. While elasticity is often confused with scalability and workload management, there are overlaps which should not be avoided \cite{elasticity-1}. Within this area, we believe that there are numerous challenges, which can be addressed for driving academic cloud research forward. 

The first challenge within elasticity is related to resource provisioning, an area which overlaps with scalability \cite{elasticity-2, elasticity-5}. Through the Infrastructure as a Service layer large pools of abstracted resources can be quickly provisioned. This facilitates a wide range of research opportunities. Notably the issues of efficient provisioning are not yet fully addressed. There is a need for research which investigates how to avoid under and over provisioning, how to provision for performance and cost, and how to efficiently compute requirements. 

The second challenge is related to workload management, which again overlaps with the issue of scalability \cite{elasticity-6}. How can frameworks use workload information for provisioning resources, or is it possible to anticipate the workload which can be used to provision resources thereby reducing the effects of variable times in provisioning resources. 

The third challenge is related to modelling elasticity \cite{elasticity-3, elasticity-4}. An attractive feature of the cloud is its abstract view of elasticity and how easily resources can be provisioned. One important question that arises is can the abstract view also present different views of the system. For example, what level of performance can be obtained by provisioning $n$ resources when the workload increases by $m\%$, what cost models can best capture these scenarios, what additional costs will be incurred when $n$ resources are added.  There are ongoing efforts to address these issues (e.g., the SPEC RG Cloud Working Group~\cite{SPEC}), but we believe this needs to be a direction for refocusing academic cloud computing research. Addressing these challenges can elicit collaborations with industry.

\section{Conclusion}
Contemplating whether academics are addressing the wrong class of problems to have any lasting impact on the cloud computing community motivated this discussion paper. Its objective is to foster a useful discussion about how academia can focus its efforts to solve meaningful problems. 

We have argued that there are five fundamental pitfalls, which restrict academics from conducting research at the infrastructure level, which is where the vast majority of current academic research focuses. These pitfalls are related to restricted access to data centre sized resources, the abstraction provided by the cloud, reproducing results in a real cloud environment, rebranding previous research, and a lack of industrial relations that prohibit meaningful cloud research in academia. 

We expect to see genuine cloud research focus less upon physical infrastructure and continue to embrace abstraction through user driven research, programming models, and PaaS clouds. In addition research needs to focus on providing software engineers with better tools to program truly elastic applications, and debug large-scale applications, which utilise existing cloud infrastructure.

{\footnotesize \bibliographystyle{acm}
\bibliography{../cloud}}


\end{document}